\documentclass[letterpaper]{article}
\usepackage{spconf}
\makeatletter
\def\bstctlcite{\@ifnextchar[{\@bstctlcite}{\@bstctlcite[@auxout]}}
\def\@bstctlcite[#1]#2{\@bsphack
  \@for\@citeb:=#2\do{%
    \edef\@citeb{\expandafter\@firstofone\@citeb}%
    \if@filesw\immediate\write\csname #1\endcsname{\string\citation{\@citeb}}\fi}%
  \@esphack}
\makeatother

\usepackage{cite}
\usepackage{amsmath,amssymb,amsfonts}
\usepackage{graphicx}
\usepackage{textcomp}
\usepackage{xcolor}
\usepackage{booktabs}
\usepackage{array}
\usepackage{pgfplots}
\usepackage[hidelinks]{hyperref}
\usepackage{orcidlink}
\pgfplotsset{compat=1.18}
\usetikzlibrary{arrows.meta,positioning}

\renewcommand{\footnotesize}{\small}
\definecolor{figblue}{HTML}{2A78D6}
\definecolor{figorange}{HTML}{EB6834}
\definecolor{figaqua}{HTML}{1BAF7A}

\begin{document}
\ninept
\bstctlcite{IEEEexample:BSTcontrol}

\title{WORD-ALIGNED PROSODY REFLECTS SPEAKER DIFFERENCES MORE THAN DEPRESSION OR PTSD SEVERITY}
\name{Hsiang-Chen Yeh\orcidlink{0009-0004-5613-4814}$^{1}$, Aurosweta Mahapatra\orcidlink{0009-0001-3057-2021}$^{2}$, Shreeram Suresh Chandra\orcidlink{0000-0002-1085-2544}$^{2}$,}
\makeatletter\gdef\@name{{\em Hsiang-Chen Yeh\orcidlink{0009-0004-5613-4814}$^{1}$, Aurosweta Mahapatra\orcidlink{0009-0001-3057-2021}$^{2}$, Shreeram Suresh Chandra\orcidlink{0000-0002-1085-2544}$^{2}$,}\\{\em Ryan L. Boyd\orcidlink{0000-0002-1876-6050}$^{3}$, Berrak Sisman\orcidlink{0000-0001-8078-3305}$^{2}$}\\}\makeatother
\address{$^{1}$Clinical Mental Health Counseling, Johns Hopkins University, Baltimore, MD, USA\\
\kern-1pt$^{2}$Center for Language and Speech Processing (CLSP), Johns Hopkins University, Baltimore, MD, USA\kern-1pt\\
$^{3}$College of Connected Computing, Vanderbilt University, Nashville, TN, USA}

\maketitle

\begin{abstract}

Word-aligned prosody is increasingly used to detect depression and posttraumatic stress disorder (PTSD), yet the stability and clinical relevance of these measurements remain unclear. We examine stability at two levels. At the word level, we ask whether a speaker produces the same word in a similar speaker-specific way across different utterance contexts. At the speaker level, we ask whether averaging prosody across many words produces consistent speaker scores across separate recordings. Drawing on principles of psychological measurement, we compare these representations in four corpora spanning acted, spontaneous, and clinical speech. Word-specific deviations show little cross-context recurrence (0.01--0.16), whereas speaker summaries are highly consistent, reaching 0.997 in ESD, 0.73--0.90 in MSP-Podcast, and 0.84--0.996 in DAIC-WOZ and PTSD-STOP. Despite this consistency, associations with depression or PTSD severity do not exceed 0.15; equivalence tests exclude effects of 0.30 in all analyses and 0.20 in DAIC-WOZ, while within-person associations with daily PTSD symptoms are near zero ($|\rho|\leq0.04$). Thus, aggregation yields stable speaker-level prosodic measures, but these measures do not detectably track depression or PTSD severity in the studied corpora. Code is available at \url{https://github.com/jen900704/word-aligned-prosody}.

\end{abstract}


\section{Introduction}
\label{sec:intro}

Depression and PTSD severity is commonly assessed through self-report, motivating interest in complementary behavioral measures. Speech is promising because it captures both what a person says and how they say it~\cite{cummins2015review}. Advances in word-level alignment now allow acoustic features to be assigned to individual words~\cite{bain2023whisperx}, and these features are increasingly used in depression and dementia classification~\cite{rohanian2019detecting,ortiz2025cognialign}. Reproducible here means that a prosodic measurement remains consistent when it is computed again from different speech produced by the same person. Whether word-level measurements are reproducible in this sense has not been established; prior speech research has examined reproducibility only at the recording or session level~\cite{stegmann2020repeatability}.

Using a word-level acoustic value as evidence of a clinical condition assumes that it reflects the speaker's current state. However, an apparently depression-related vocal pattern may instead reflect the speaker's baseline voice, recording environment, or microphone, entangling condition-related information with speaker identity~\cite{yeh2026speaking}. We therefore ask how much of a word-level acoustic value is a stable speaker property and how much is specific to the occasion on which the word is spoken. This distinction determines whether word-level prosody can provide reliable evidence of depression or PTSD.

Personality psychology distinguishes between stable person-level properties and behavior specific to an occasion. A word-level prosodic value can similarly be viewed as a single behavioral observation. Mischel found that behavior in one situation predicted behavior in another at correlations below 0.3, termed ``personality coefficients,'' corresponding to less than 10\% shared variance~\cite{mischel1968book}. The resulting person--situation literature concluded that individual observations have low but nonzero stability~\cite{mischel1995cognitive}. Subsequent daily-diary research showed that this limitation concerns the unit of measurement rather than the stability of the person~\cite{epstein1979stability}. Although one day's behavior predicted another at only 0.2--0.3, averages across odd and even days correlated at 0.8--0.9. A trait can therefore be understood as a distribution of states, with each occasion representing one observation from that distribution~\cite{fleeson2001toward}. Aggregation has produced similar gains in language-based prediction: county-level income prediction improved from 0.73 to 0.82 when tweets were first summarized by user, preventing highly active users from dominating the estimate~\cite{giorgi2018remarkable}. We apply the same principle by aggregating words within speakers while holding the corpus, features, and extraction procedure fixed. This approach also reflects clinical measurement: depression and PTSD are defined over periods of at least two weeks and one month, respectively~\cite{apa2013dsm}, as reflected by the assessment windows of the Patient Health Questionnaire (PHQ-8) and the PTSD Checklist (PCL)~\cite{kroenke2009phq,blevins2015posttraumatic}.

A word could nevertheless function as a recurring situation that evokes a stable, speaker-specific response. In prior work, individuals exhibited stable behavioral profiles across recurring situations after their overall behavioral levels were removed, producing correlations of 0.47 for verbal aggression and 0.41 for compliance~\cite{mischel1995cognitive}. A speaker-by-word interaction represents an analogous \emph{if--then} signature: speakers may differ consistently in how they realize particular words. This yields two interpretations of word-level prosody. If each word token is primarily an occasion, word-specific deviations should recur poorly while summaries across words remain stable. If the word acts as a recurring situation, its speaker-specific deviation should recur across contexts. We test both interpretations and then examine whether the reproducible quantity is associated with symptoms.

We address two research questions. \textbf{RQ1:} Do speaker-specific realizations of individual words recur across contexts, and does aggregation across words yield reproducible speaker summaries? \textbf{RQ2:} Are these summaries associated with depression (PHQ-8) or PTSD severity (PCL), either between or within persons? We examine four corpora spanning controlled, spontaneous, and clinical speech: ESD~\cite{zhou2021seen,zhou2022emotional}, MSP-Podcast~\cite{lotfian2017building,busso2025msp}, DAIC-WOZ~\cite{gratch2014distress}, and PTSD-STOP~\cite{ringwald2025day}. Using generalizability theory, we partition prosodic variance into speaker, word, speaker-by-word, and residual components~\cite{vanleeuwen1998generalizability}. Following the V3 framework~\cite{goldsack2020verification}, we treat reproducibility as a first component of analytical validation before testing clinical validity through associations with symptoms. We find that word-aligned prosody primarily captures an additive speaker level, with little stable speaker-by-word structure. When aggregated into speaker summaries, the same ESD recordings achieve a reproducibility of 0.997, although sex and recording channel partly contribute to this ceiling. Despite their reproducibility, the summaries show no detectable association with symptom severity. 

To our knowledge, this is the first study to evaluate the cross-context reproducibility of word-level prosodic features as potential behavioral markers of depression and PTSD.

\section{Study Design}
This section describes the design used to test whether per-word deviations recur, whether speaker summaries are reproducible, and whether such summaries relate to symptom severity. We introduce the corpora and their shared features, then the analysis order. Only speech attributed to the target speaker or participant is used. \emph{Context} throughout means the utterance a word occurs in, e.g.\ a sentence frame within an emotion condition in ESD. ESD and MSP-Podcast address RQ1; DAIC-WOZ and PTSD-STOP address RQ2 after their own reproducibility checks.
\subsection{Corpora}

\subsubsection{ESD: acted emotional speech}
ESD's ten English speakers are gender balanced~\cite{zhou2022emotional}; we use nine (0014 excluded for inaccessible transcripts), four male and five female, who each produce the same 350 sentences in five emotion conditions~\cite{zhou2021seen,zhou2022emotional}. Words recur across sentence contexts, so recurrence is estimable within condition. ESD is acted, carries no clinical label, and its emotion labels are not analyzed. Acted speech is used because it induces pronounced changes in prosody while holding the speaker and the words fixed, a controlled test of whether speaker-specific word patterns survive a change of speaking style. 

\subsubsection{MSP-Podcast: Spontaneous Emotional Speech}

MSP-Podcast provides spontaneous speech from multiple recordings per speaker, making it possible to evaluate reproducibility across independent recording contexts~\cite{lotfian2017building,busso2025msp}. A prespecified selection requiring at least 100 aligned utterances and recordings from at least two episodes retained 99 identified speakers (62 male and 37 female; 551,759 word tokens). We assigned complete episodes, rather than individual utterances, to each split half. Thus, the two halves for a speaker never contained speech from the same recording session. Emotion labels were not used.

\subsubsection{DAIC-WOZ: clinical interview}
DAIC-WOZ contains semi-structured clinical interviews in English, conducted by a virtual interviewer controlled by a human operator and designed to elicit signs of psychological distress~\cite{gratch2014distress}. We use the 189 participants with a PHQ-8 score, 56 of whom carry the corpus's binary depression label; the score enters our analyses as a continuous measure. Interviewer prompts are excluded because they leak diagnostic information~\cite{burdisso2024daic}. DAIC-WOZ is the corpus on which word-level depression classifiers, including the one that motivated this study, were developed~\cite{rohanian2019detecting}, so it is where a per-word reading of prosody would be applied.

\subsubsection{PTSD-STOP: daily video diaries}
PTSD-STOP follows World Trade Center responders recruited through the Stony Brook World Trade Center Health Program~\cite{ringwald2025day,ganesan2026word}. For up to 90 days, each participant recorded a daily video diary about their day on a personal smart device and rated their PTSD symptoms the same day. We use 13{,}600 daily monologues from 219 responders (84\% male). The outcome is a daily-adapted PCL score: eight PCL items and an overall stress rating on five-point scales, giving 1--5~\cite{ganesan2026word}. Its mean over days 0--60 exists for 213 participants, and repeated same-day scores for 178. PTSD-STOP supplies what the other corpora cannot: many occasions per person, each with a symptom rating from the same day, so prosody can be tested against symptoms between persons, across change, and within persons over days.

\subsection{Shared word-aligned features}
All corpora share four feature definitions: mean F0, robust F0 range (P90--P10), energy in dB, and syllable-adjusted log word duration.

\textbf{Extraction.} Reference transcripts of ESD, DAIC-WOZ and PTSD-STOP are aligned to 16-kHz mono audio with WhisperX 3.3.1~\cite{bain2023whisperx} (alignment model in the protocol). MSP-Podcast uses the forced alignments distributed with the corpus. Within word boundaries, F0 is estimated by autocorrelation from voiced frames only (parameters in the protocol). Energy is $20\log_{10}$ of word RMS, and syllables come from CMUdict. The same feature code is used in all four corpora.

\textbf{Analysis order.} Thresholds, feature order and failed-fit rules were fixed before any estimate was inspected. Symptom scores were withheld until the corresponding feature had passed its reproducibility checks. Analyses added after viewing results are labeled post hoc, and none altered a prespecified decision. 

\begin{table}[!t]
\centering
\caption{RQ1 evidence. $\bar R_c$: ESD cross-context recurrence of the speaker-by-word deviation over five conditions (9 speakers, 98{,}202 tokens). $r_c$: MSP-Podcast corrected split-half consistency of the speaker summary (99 speakers). Brackets: bootstrap intervals, biased low with nine ESD clusters (Sec.~\ref{sec:esdrc}). n/a: no interval for duration (576/1{,}000 fits). G: gate; undet.\ = a robustness fit failed. Zeros omitted.}
\label{tab:evidence}
\small
\setlength{\tabcolsep}{3pt}
\begin{tabular}{lcccc}
\toprule
& \multicolumn{2}{c}{ESD word deviation} & \multicolumn{2}{c}{MSP speaker summary} \\
\cmidrule(lr){2-3}\cmidrule(lr){4-5}
Feature & $\bar R_c$ [CI] & G & $r_c$ [CI] & G \\
\midrule
Mean F0 & .059 [.021, .059] & fail & .895 [.816, .947] & pass \\
F0 range & .160 [.075, .163] & undet. & .732 [.494, .881] & pass \\
Energy & .029 [.013, .029] & fail & .869 [.712, .935] & pass \\
Log duration & .010 [n/a] & undet. & .793 [.688, .864] & fail \\
\bottomrule
\end{tabular}
\end{table}

\section{RQ1: Is Word-Aligned Prosody Stable Across Contexts and Recordings?}
RQ1 evaluates two distinct forms of stability. First, we test whether a speaker's distinctive realization of a particular word recurs across utterance contexts. We define this \emph{speaker-specific word deviation} relative to both the speaker's overall prosodic level and the word's typical realization. Second, we test whether aggregating measurements across many words yields a \emph{speaker summary} that remains consistent across separate recordings. Distinguishing these estimands reveals whether prosodic stability is specific to individual words or emerges through aggregation at the speaker level.

We interpret a feature only if its recurrence exceeds 0.70 for ESD word deviations (Gate E) or its consistency exceeds 0.60 for speaker summaries (Gate M). Each feature must also satisfy the prespecified bootstrap, leave-one-speaker-out, and equal-weight robustness checks described in the protocol.

\subsection{Cross-context recurrence in ESD}
To estimate how far a speaker's word deviation recurs, we separate speaker ($S$), word ($W$) and speaker-by-word ($S{\times}W$, subscript $SW$) variation within each emotion condition $c$:
\begin{equation}
 y_{swr}^{(c)} = \mu_c + u_{S,s}^{(c)} + u_{W,w}^{(c)} + u_{SW,sw}^{(c)} + \varepsilon_{swr}^{(c)},
\end{equation}
where $s$ is the speaker, $w$ the lexical item and $r$ an occurrence. The residual includes changes across utterance contexts. A post hoc variant adds an utterance random effect so that shared sentence context leaves the residual. Cross-context recurrence is $R_c=\sigma^2_{SW,c}/(\sigma^2_{SW,c}+\sigma^2_{e,c})$, the single-observation intraclass correlation of the interaction term: the share of the $SW$ deviation that recurs, not the recurrence of the raw word observation. Between-speaker differences, sex included, enter $\sigma^2_S$. We report the mean over five conditions, $\bar R_c=\frac{1}{5}\sum_c R_c$. Mixed-model (REML) fits use cells with at least two finite observations. For duration, syllable and speaker fixed effects are removed first. Syllable count mechanically sets duration, and speech rate would otherwise dominate $\sigma^2_S$. Duration's $S$ is therefore a residual, not between-speaker variance, and is not comparable to the other three in Fig.~\ref{fig:varcomp}; the summary analysis of Sec.~\ref{sec:esdmsp} keeps speaker differences instead. 

A raw token, which retains the speaker's level, carries $(\sigma^2_S+\sigma^2_{SW})/(\sigma^2_S+\sigma^2_{SW}+\sigma^2_e)=0.23$--$0.37$ across those three features. This is the speaker's share of one observation, an acoustic analogue of the personality coefficient. Because that information sits in $\sigma^2_S$, averaging over words recovers it, and a speaker summary can be highly consistent even when its word deviations do not recur (Fig.~\ref{fig:agg}).

\subsection{Consistency across recordings in MSP-Podcast}
\label{sec:msp}
Word values are adjusted for relative utterance position $p$ ($y=\beta_0+\beta_1p+\varepsilon$; duration also for syllable count $\ell$), fit once without split labels; the residuals retain between-speaker differences. Each speaker's two recording halves are summarized by the median adjusted word value. Summaries are correlated (Pearson) across speakers with at least 50 valid observations per half, then Spearman--Brown corrected, $2r/(1+r)$. The correction is an extrapolation for medians and unequal halves; it concerns consistency, not absolute agreement~\cite{koo2016guideline}. The sample and 25 alternative splits were fixed before processing audio.

\section{RQ2: Does Speaker-Level Prosody Track Depression and PTSD Severity?}

RQ2 applies the speaker-summary procedure from Sec.~\ref{sec:msp} to the clinical corpora, DAIC-WOZ and PTSD-STOP. We first evaluate the reproducibility of each prosodic feature and then test its association with symptom severity. In PTSD-STOP, split halves are constructed from separate sets of complete recordings, providing consistency across recording occasions. DAIC-WOZ contains only one interview per participant; its odd--even split therefore measures consistency within an interview rather than stability across sessions.

We assess clinical associations at three levels. First, we compute between-person Spearman correlations between speaker summaries and PHQ-8 scores in DAIC-WOZ or mean PCL scores in PTSD-STOP. Second, for PTSD-STOP, we calculate changes in prosody and PCL between each participant's earliest and latest recording sets and correlate the two change scores. We apply Benjamini--Hochberg correction separately to each prespecified family of four feature tests: DAIC-WOZ between-person, PTSD-STOP between-person, and PTSD-STOP change.

Finally, motivated by the occasion-level variation observed in RQ1, we conduct a post hoc within-person analysis in PTSD-STOP. This analysis tests whether daily prosody, expressed as a deviation from each participant's mean, is associated with same-day PCL severity. The model includes a random intercept for each participant, with a random-slope specification used as a robustness check.

\section{Results}
\subsection{RQ1: Two scores from the same ESD audio}
\label{sec:esdmsp}
To compare the two units on identical data, a prespecified analysis applied the MSP speaker-summary procedure to ESD itself. Audio, speakers, features and extraction were unchanged, and halves were formed from whole prompts. Corrected consistency ranged 0.9973--0.9996, though with four male and five female speakers part of that ceiling is sex and channel, not speaker. The same recordings scored as word deviations recur below 0.16 (Table~\ref{tab:evidence}). This is the contrast of Sec.~\ref{sec:intro} between one occasion and an aggregate, with the audio held constant.

\subsection{RQ1: Strong speaker level, weak word signature}
\label{sec:esdrc}
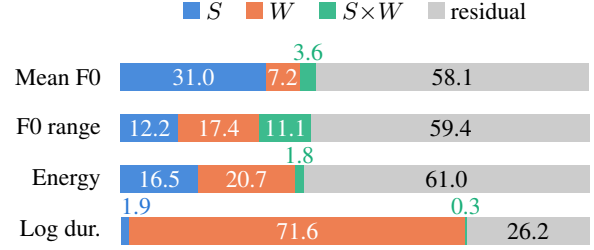
\begin{figure}[!t]
\centering
\begin{tikzpicture}[x=1mm,y=1mm]
\tikzset{lab/.style={font=\fontsize{9}{9}\selectfont,inner sep=0pt}}
\node[anchor=south,font=\small] at (31.0,1.2) {\tikz\fill[figblue!85] (0,0) rectangle (2.2mm,2.2mm); $S$\quad \tikz\fill[figorange!85] (0,0) rectangle (2.2mm,2.2mm); $W$\quad \tikz\fill[figaqua!85] (0,0) rectangle (2.2mm,2.2mm); $S{\times}W$\quad \tikz\fill[black!20] (0,0) rectangle (2.2mm,2.2mm); residual};
\node[anchor=east,font=\small] at (-1.5,-5.05) {Mean F0};
\fill[figblue!85] (0.000,-6.85) rectangle (19.220,-3.25);
\node[lab,text=white] at (9.610,-5.05) {31.0};
\fill[figorange!85] (19.220,-6.85) rectangle (23.684,-3.25);
\node[lab,text=white] at (21.452,-5.05) {7.2};
\fill[figaqua!85] (23.684,-6.85) rectangle (25.916,-3.25);
\node[lab,text=figaqua,anchor=south] at (24.800,-2.95) {3.6};
\fill[black!20] (25.916,-6.85) rectangle (61.938,-3.25);
\node[lab,text=black] at (43.927,-5.05) {58.1};
\node[anchor=east,font=\small] at (-1.5,-11.85) {F0 range};
\fill[figblue!85] (0.000,-13.65) rectangle (7.564,-10.05);
\node[lab,text=white] at (3.782,-11.85) {12.2};
\fill[figorange!85] (7.564,-13.65) rectangle (18.352,-10.05);
\node[lab,text=white] at (12.958,-11.85) {17.4};
\fill[figaqua!85] (18.352,-13.65) rectangle (25.234,-10.05);
\node[lab,text=white] at (21.793,-11.85) {11.1};
\fill[black!20] (25.234,-13.65) rectangle (62.062,-10.05);
\node[lab,text=black] at (43.648,-11.85) {59.4};
\node[anchor=east,font=\small] at (-1.5,-18.65) {Energy};
\fill[figblue!85] (0.000,-20.45) rectangle (10.230,-16.85);
\node[lab,text=white] at (5.115,-18.65) {16.5};
\fill[figorange!85] (10.230,-20.45) rectangle (23.064,-16.85);
\node[lab,text=white] at (16.647,-18.65) {20.7};
\fill[figaqua!85] (23.064,-20.45) rectangle (24.180,-16.85);
\node[lab,text=figaqua,anchor=south] at (23.622,-16.55) {1.8};
\fill[black!20] (24.180,-20.45) rectangle (62.000,-16.85);
\node[lab,text=black] at (43.090,-18.65) {61.0};
\node[anchor=east,font=\small] at (-1.5,-25.45) {Log dur.};
\fill[figblue!85] (0.000,-27.25) rectangle (1.178,-23.65);
\node[lab,text=figblue,anchor=south west] at (0.000,-23.35) {1.9};
\fill[figorange!85] (1.178,-27.25) rectangle (45.570,-23.65);
\node[lab,text=white] at (23.374,-25.45) {71.6};
\fill[figaqua!85] (45.570,-27.25) rectangle (45.756,-23.65);
\node[lab,text=figaqua,anchor=south] at (45.663,-23.35) {0.3};
\fill[black!20] (45.756,-27.25) rectangle (62.000,-23.65);
\node[lab,text=black] at (53.878,-25.45) {26.2};
\end{tikzpicture}
\caption{ESD variance decomposition, averaged over the five conditions, in \% of total variance: speaker ($S$), word ($W$), their interaction ($S{\times}W$, the candidate word-specific signature) and residual. $R_c=S{\times}W/(S{\times}W+\text{residual})$. Duration's $S$ is a residual (syllable and speaker effects removed first), not between-speaker variance.}
\label{fig:varcomp}
\end{figure}
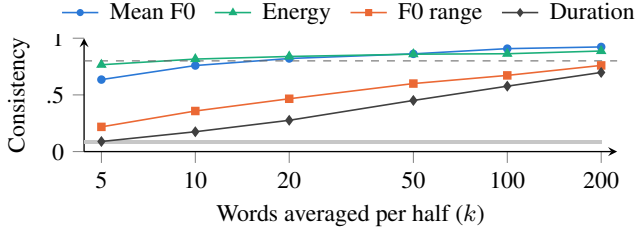
\begin{figure}[!t]
\centering
\begin{tikzpicture}
\begin{axis}[
  scale only axis, width=0.82\linewidth, height=15mm,
  xmode=log, log basis x=10,
  xmin=4.4, xmax=225, ymin=0, ymax=1,
  xtick={5,10,20,50,100,200}, xticklabels={5,10,20,50,100,200},
  ytick={0,0.5,1}, yticklabels={0,.5,1},
  xlabel={Words averaged per half ($k$)},
  ylabel={Consistency},
  label style={font=\small}, tick label style={font=\small},
  xlabel near ticks, ylabel near ticks,
  axis lines=left, tick align=outside, axis line style={line width=0.5pt},
  legend style={at={(0.5,1.04)},anchor=south,legend columns=4,draw=none,font=\small,
                /tikz/every even column/.append style={column sep=5pt},
                /tikz/every odd column/.append style={column sep=2pt}},
  legend image post style={scale=0.8},
]
\fill[black!22] (axis cs:4.4,0.067) rectangle (axis cs:200,0.095);
\addplot[color=figblue,mark=*,mark size=1.3pt,semithick] coordinates {(5,.635)(10,.759)(20,.821)(50,.862)(100,.908)(200,.923)};
\addlegendentry{Mean F0}
\addplot[color=figaqua,mark=triangle*,mark size=1.7pt,semithick] coordinates {(5,.766)(10,.816)(20,.839)(50,.859)(100,.863)(200,.887)};
\addlegendentry{Energy}
\addplot[color=figorange,mark=square*,mark size=1.3pt,semithick] coordinates {(5,.218)(10,.358)(20,.466)(50,.600)(100,.672)(200,.760)};
\addlegendentry{F0 range}
\addplot[color=black!75,mark=diamond*,mark size=1.6pt,semithick] coordinates {(5,.089)(10,.175)(20,.276)(50,.451)(100,.577)(200,.698)};
\addlegendentry{Duration}
\draw[dashed,black!60] (axis cs:4.4,0.8) -- (axis cs:225,0.8);
\end{axis}
\end{tikzpicture}
\caption{Aggregation in MSP-Podcast, one corpus, two estimands. Lines: observed split-half consistency of a speaker summary from $k$ sampled words per half (200 draws, seed 13). Gray band: word-deviation recurrence, nine-speaker draws (post hoc; .07--.10); averaging different words cannot raise it. Dashed: 0.80.}
\label{fig:agg}
\end{figure}

Fig.~\ref{fig:varcomp} illustrates that the $S{\times}W$ share is small everywhere (0.3\%--11.1\%, against a residual of 26--61\%). The recurrence of all four features falls far below Gate E's 0.70 (Table~\ref{tab:evidence}). Mean F0 and energy fail outright; F0 range and duration are undetermined. ESD has only nine speakers. Rerunning the same estimator on 20 random draws of nine MSP-Podcast speakers (post hoc, matching the ESD fit size) gave median recurrence 0.067--0.095, with no draw above 0.163 (Fig.~\ref{fig:agg}).

Four post hoc extraction checks (described in the protocol) leave $\bar R_c$ below 0.25; the utterance effect absorbed 0.3--19\% of variance yet moved it by at most 0.012.

The shortfall is not explained by measurement noise alone. The speaker-by-word component is itself small (duration excluded, its speaker effects being removed first), and no extraction check enlarges it. The utterance effect shows much of the residual to be shared utterance-level structure rather than noise. This pattern is the reliability paradox described in~\cite{hedge2018reliability}. Words differ robustly ($W$ is 7--72\% of variance) while speakers' deviations from them do not, so little stable word-specific deviation remains for a second context to reproduce.

\subsection{RQ1: Speaker summaries and the unit of aggregation}
\label{sec:mspres}
Over 551{,}759 tokens from 99 speakers, MSP split-half consistency was 0.732--0.895 corrected. Mean F0, F0 range and energy passed Gate M. Duration did not. Its token-weighted consistency of 0.793 fell to 0.560 under equal recording weight, while the other three moved by at most 0.038. This is the super-user effect described in~\cite{giorgi2018remarkable}. Duration covaries with how much a speaker talks, so it is the feature most exposed to a summary that weights observations rather than persons. As a practical rule (Fig.~\ref{fig:agg}), mean F0 and energy reach 0.80 within 10--20 words; F0 range and duration do not by 200.

\subsection{RQ2: Analytical validity without detectable clinical validity}
\label{sec:clinres}
In V3 terms, analytical validation came first. Both clinical corpora passed their gates on all four features before any symptom score was seen (PTSD-STOP 0.975--0.980, DAIC-WOZ 0.844--0.996). Low reliability of the acoustic summary is therefore unlikely to account for a clinical null.

\begin{table}[!t]
\centering
\caption{PTSD-STOP associations: Spearman $\rho$ with participant bootstrap intervals. Benjamini--Hochberg within each family, neither surviving: between-person ($N=213$) $q\ge.109$, change ($N=178$) $q\ge.295$. DAIC-WOZ ($N=189$) is null throughout ($|\rho|\le.044$, bounds within $[-.20,.19]$).}
\label{tab:clinical}
\small
\setlength{\tabcolsep}{4pt}
\begin{tabular}{lcc}
\toprule
Feature & between-person $\rho$ [CI] & change $\rho$ [CI] \\
\midrule
Mean F0 & .151 [.013, .283] & .047 [$-$.110, .194] \\
F0 range & .032 [$-$.112, .169] & .110 [$-$.034, .249] \\
Energy & .051 [$-$.076, .186] & $-$.109 [$-$.259, .045] \\
Log duration & .056 [$-$.072, .180] & $-$.013 [$-$.162, .135] \\
\bottomrule
\end{tabular}
\end{table}

No between-person correlation with PCL or PHQ-8 survived correction (Table~\ref{tab:clinical}). Post hoc equivalence tests (two one-sided tests)\footnote{On Spearman correlations, Fisher-$z$ with the Bonett--Wright standard error, validated against the study's bootstrap; the within-person family uses its clustered bootstrap directly.} bound each family: the smallest association excluded is 0.16 for DAIC-WOZ, 0.26 for PTSD-STOP between persons and 0.23 for change. In individual-differences research 0.20 is a typical effect and 0.30 a large one~\cite{funder2019evaluating,gignac2016effect}, and single-feature correlations with severity in longitudinal depression samples run about 0.1--0.3~\cite{trevino2011phonologically,quatieri2012vocal,horwitz2013relative}. All three families exclude 0.30; only DAIC-WOZ excludes 0.20, so PTSD-STOP remains compatible with effects of the size that the literature reports. In DAIC-WOZ, where 56 of 189 participants carry the depression label, a small difference confined to that group ($d=0.3$, $\rho\approx0.14$) would also fall inside the bound. Mean F0 in PTSD-STOP was the largest and the only interval excluding zero. It fails correction, and with no sex variable in our extract, a sex-related F0 difference cannot be ruled out.

The post hoc within-person model (5{,}311 participant-days, median 29) tests whether daily prosody tracks daily symptoms, and finds that it does not. Within-person Spearman correlations were 0.022, 0.037, $-0.031$ and 0.016 (cluster-bootstrap intervals within $[-0.06,0.07]$, equivalent to zero against 0.07). No slope survived once slopes varied by person.

\section{Discussion}
Word deviations have low reliability, but the variance they leave behind belongs largely to the occasion rather than to noise. Speaker summaries have near-ceiling reliability and no detectable validity against symptom scales.

If--then signatures reproduce when situations are defined by psychologically active features, which is that theory's own boundary condition~\cite[p.~250]{mischel1995cognitive}, and lexical identity is not such a feature for prosody. A word is not a situation.

For the systems that motivated this study~\cite{rohanian2019detecting,ortiz2025cognialign}, reading the per-word sequence in context is compatible with our result; reading a single word's value as an attribution about the speaker is not. And $\sigma^2_S$ is partly channel (MSP-Podcast splits cross episodes). Under speaker- or episode-overlapping splits, per-word streams therefore risk identity and device leakage~\cite{yeh2026speaking}. A model that seems to learn how this person says this word may be learning which microphone this is.

\textbf{Limitations.} Four English corpora and four suprasegmental features do not establish generality across languages, segmental features or learned representations. PTSD-STOP's daily-adapted PCL has no published reliability. Post hoc, on our 213 participants its single-day ICC is 0.56 (within-person SD 0.58, between 0.76 on the 1--5 scale). The 60-day mean is therefore reliable near 0.97, so the between-person null is not attenuation. Whether day-to-day variance is change or error is unquantified, though language tracks it within persons ($r=0.21$--$0.30$~\cite{ganesan2026word}).

\section{Conclusion}
A word is not a situation. The same audio recurs below 0.16 as word deviations and reaches 0.997 as a speaker summary, a ceiling partly due to sex and channel. Speaker information appears once values are aggregated within a speaker; aligning them to words adds little that recurs. For systems that read per-word prosody, this argues for aggregating within a speaker before interpreting, and for testing any word-level claim against the occasion it came from. Symptom severity was not detectable in either clinical corpus, whether between persons, across change, or within persons over days. What unit, if any, links prosody to symptoms remains open.

\section{Compliance with Ethical Standards}
ESD, MSP-Podcast and DAIC-WOZ were analyzed under their dataset-use terms; PTSD-STOP under written PI permission and IRB protocol addition, with all analysis on the host laboratory's servers. The protocol and amendment log are available from the authors on request; code is available at \url{https://github.com/jen900704/word-aligned-prosody}; PTSD-STOP audio and outcomes cannot be shared.

\let\OLDthebibliography\thebibliography
\renewcommand{\thebibliography}[1]{\OLDthebibliography{#1}\footnotesize\setlength{\itemsep}{0pt}\setlength{\parsep}{0pt}}
\bibliographystyle{IEEEtran}
\bibliography{references}

\end{document}